\documentclass[12pt]{article}
\usepackage{epsf}

\renewcommand{\bar}[1]{\overline{#1}}
\newcommand{\VEV}[1]{\left\langle{#1}\right\rangle}
\newcommand{\etal}{{\em et al.}}
\newcommand{\ie}{{\em i.e.}}
\newcommand{\eg}{{\em e.g.}}

\newcommand{\ket}[1]{\vert\,{#1}\rangle}
\newcommand{\gsim} {\buildrel > \over {_\sim}}
\newcommand{\M}{\mathcal{M}}

\begin{document}

\begin{flushright}
SLAC-PUB-9281\\
July 2002
\end{flushright}

\bigskip\bigskip

\begin{center}
{{\bf\Large Perspectives on Exclusive Processes in
QCD\footnote{Work supported by the Department of Energy, contract
DE--AC03--76SF00515.}}}

\bigskip
Stanley J. Brodsky \\
Stanford Linear Accelerator Center, Stanford University \\
Stanford, California 94309\\
e-mail: sjbth@slac.stanford.edu

\end{center}

\vfill

\begin{center}
Abstract
\end{center}
Hard hadronic exclusive processes are now at the forefront of QCD
studies, particularly because of their role in the interpretation
of exclusive hadronic $B$ decays, processes which are essential
for determining the CKM phases and the physics of $CP$ violation.
Perturbative QCD and its factorization properties at high momentum
transfer provide an essential guide to the phenomenology of
exclusive amplitudes at large momentum transfer---the leading
power fall-off of form factors and fixed-angle cross sections, the
dominant helicity structures, and their color transparency
properties. The reduced amplitude formalism provides an extension
of the perturbative QCD predictions to exclusive nuclear
amplitudes. The hard scattering subprocess $T_H$ controlling the
leading-twist amplitude is only evaluated in the QCD perturbative
domain where the propagator virtualities are above the separation
scale. A critical question is the momentum transfer required such
that leading-twist perturbative QCD contributions dominate. I
review some of the contentious theoretical issues and empirical
challenges to Perturbative QCD based  analyses, such as the
magnitude of the leading-twist contributions, the role of soft and
higher twist QCD mechanisms, the effects of non-zero orbital
angular momentum, the possibility of single-spin asymmetries in
deeply virtual Compton scattering, the role of hidden color in
nuclear wavefunctions, the behavior of the ratio of Pauli and
Dirac nucleon form factors, the origin of anomalous $J/\psi$
decays, the apparent breakdown of color transparency in
quasi-elastic proton-proton scattering, and the measurement of
hadron and photon wavefunctions in diffractive dijet production.

\vfill
\begin{center}
Invited Talk, presented at the\\
Workshop on Exclusive Processes at High Momentum Transfer\\
Jefferson Lab, Newport News, Virginia\\
May 15--18, 2002
\end{center}

\newpage

\section{Introduction}

\begin{figure}[htbp]
\begin{center}
\leavevmode {\epsfxsize=4.5in\epsfbox{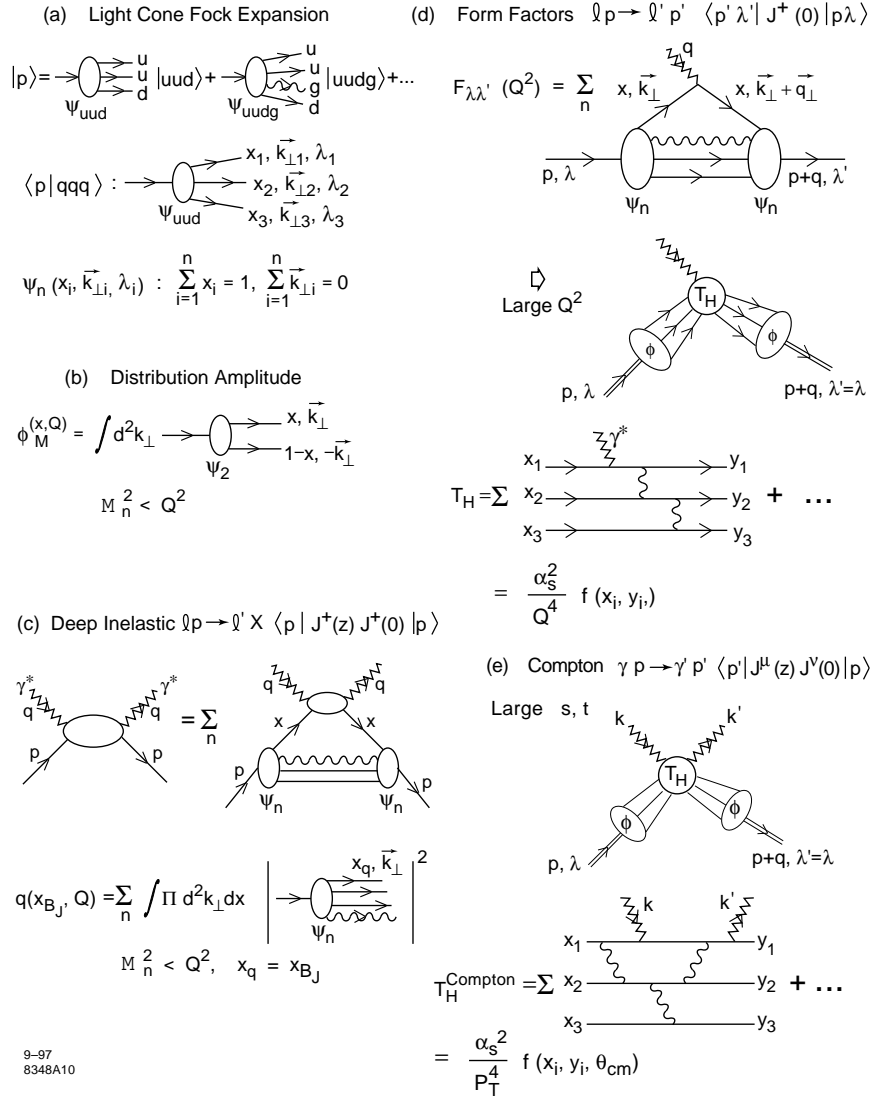}}
\end{center}
\caption[*]{ Representation of QCD hadronic processes in the
light-cone Fock expansion.  (a) The valence $uud$  and $uudg$
contributions to the light-cone Fock expansion for the proton. (b)
The distribution amplitude $\phi(x,Q)$ of a meson expressed as an
integral over its valence light-cone wavefunction restricted to $q
\bar q$ invariant mass less than $Q$.  (c) Representation of deep
inelastic scattering and the quark distributions $q(x,Q)$  as
probabilistic measures of the light-cone Fock wavefunctions.  The
sum is over the Fock states with invariant mass less than $Q$. (d)
Exact representation of spacelike form factors of the proton in
the light-cone Fock basis.  The sum is over all Fock components.
At large momentum transfer the leading-twist contribution
factorizes as the product of the hard scattering amplitude $T_H$
for the scattering of the valence quarks collinear with the
initial to final direction convoluted with the proton distribution
amplitude.  (e) Leading-twist factorization of the Compton
amplitude at large momentum transfer. \label{Fig:repc1}}
\end{figure}

\begin{figure}[htbp]
\begin{center}
\leavevmode {\epsfxsize=5in\epsfbox{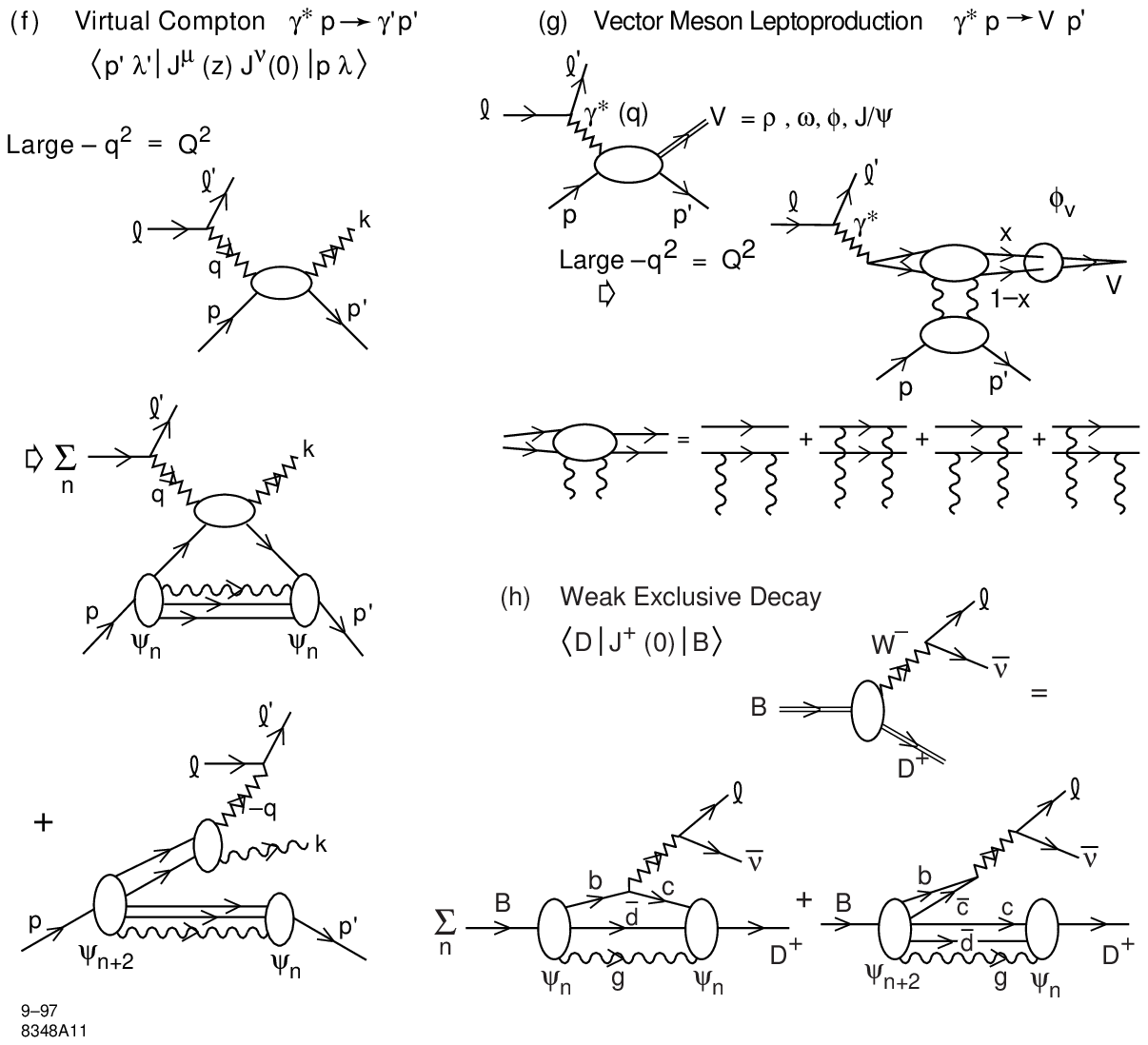}}
\end{center}
\caption[*]{ (f) Representation of deeply virtual Compton
scattering in the light-cone Fock expansion at leading twist. Both
diagonal $n  \to n$ and off-diagonal $n+2 \to n$ contributions are
required.  (g) Diffractive vector meson production at large photon
virtuality $Q^2$ and longitudinal polarization.  The high energy
behavior involves two gluons in the $t$ channel coupling to the
compact color dipole structure of the upper vertex.  The
bound-state structure of the vector meson enters through its
distribution amplitude.  (h) Exact representation of the weak
semileptonic decays of heavy hadrons in the light-cone Fock
expansion.   Both diagonal $n  \to n$ and off-diagonal pair
annihilation $n+2 \to n$ contributions are required.
\label{Fig:repc2}}
\end{figure}

Exclusive processes provide a unique window for viewing QCD
processes and hadron dynamics at the amplitude
level\cite{Brodsky:2000dr}. Hadronic exclusive processes are
closely related to exclusive hadronic $B$ decays, processes which
are essential for determining the CKM phases and the physics of
$CP$ violation. The universal light-front wavefunctions which
control hard exclusive processes such as form factors, deeply
virtual Compton scattering, high momentum transfer
photoproduction, and two-photon processes, are also required for
computing exclusive heavy hadron
decays\cite{Beneke:2000ry,Keum:2000wi,Szczepaniak:1990dt,Brodsky:2001jw},
such as $B \to K \pi$, $B \to \ell \nu \pi$, and $B \to K  p \bar
p$ \cite{Chua:2002wn}.   The same physics issues, including color
transparency, hadron helicity rules, and the question of dominance
of leading-twist perturbative QCD mechanisms enter in both realms
of physics. New tests of theory and comprehensive measurements of
hard exclusive amplitudes can be carried out for electroproduction
at Jefferson Laboratory and in two-photon collisions at CLEO,
Belle, and BaBar\cite{Brodsky:2001hv}. The perturbative QCD
approach to exclusive processes is now facing a number of strong
empirical challenges.  New data from Jefferson
Laboratory\cite{Jones:uu}  for the ratio of Pauli and Dirac form
factors of the proton appears to be at variance with QCD
expectations.  This has led to a new focus on the range of
validity of leading-twist perturbative QCD predictions and the
necessity to have better theoretical control on higher-twist
contributions.  The Pauli form factor is particularly interesting,
since it measures spin-orbit $\vec S \cdot \vec L$ couplings and
thus the presence of orbital angular momentum in the proton
light-front wavefunction. The new Jefferson Laboratory results
appear to call into question hadron helicity
conservation\cite{Brodsky:1981kj,Chernyak:1999cj}, a key feature
of the leading-twist predictions.  It is often claimed that the
leading-twist predictions for the spacelike pion and proton form
factors strongly underestimate their empirical magnitudes.  The
assumed relation between diffractive dijet production to the shape
of the projectile light-front wavefunction has also been
questioned. I will give my perspective on these challenges to
theory in this report. QCD mechanisms for exclusive processes are
illustrated in Figs.~\ref{Fig:repc1} and \ref{Fig:repc2}.

\vspace{.5cm}

\vspace{.5cm}

\section{Perturbative QCD and Exclusive Processes}

There has been considerable progress analyzing exclusive and
diffractive reactions at large momentum transfer from first
principles in QCD. Rigorous statements can be made on the basis of
asymptotic freedom and factorization theorems which separate the
underlying hard quark and gluon subprocess amplitude from the
nonperturbative physics of the hadronic wavefunctions.  The
leading-power contribution to exclusive hadronic amplitudes such
as quarkonium decay, heavy hadron decay,  and scattering
amplitudes where hadrons are scattered with large momentum
transfer can often be factorized as a  convolution of distribution
amplitudes $\phi_H(x_i,\Lambda)$ and hard-scattering quark/gluon
scattering amplitudes $T_H$ integrated over the light-cone
momentum fractions of the valence quarks\cite{Lepage:1980fj}:
\begin{eqnarray}
\M_{\rm Hadron} &=&\int
 \prod \phi_H^{(\Lambda)} (x_i,\lambda_i)\, T_H^{(\Lambda)} dx_i\ .
\label{eq:e}
\end{eqnarray}
Here $T_H^{(\Lambda)}$ is the underlying quark-gluon subprocess
scattering amplitude in which each incident and final hadron is
replaced by valence quarks with collinear momenta $k^+_i =x_i
p^+_H$, $\vec k_{\perp i} = x_i \vec p_{\perp H }.$ The invariant
mass of all intermediate states in $T_H$ is evaluated above the
separation scale $\M^2_n > \Lambda^2$. The essential part of the
hadronic wavefunction is the distribution amplitude
\cite{Lepage:1980fj}, defined as the integral over transverse
momenta of the valence (lowest particle number) Fock wavefunction;
\eg\ for the pion
\begin{equation}
\phi_\pi (x_i,Q) \equiv \int d^2k_\perp\, \psi^{(Q)}_{q\bar q/\pi}
(x_i, \vec k_{\perp i},\lambda) \label{eq:f}
\end{equation}
where the separation scale $\Lambda$ can be taken to be order of
the characteristic momentum transfer $Q$ in the process. It should
be emphasized that the hard scattering amplitude $T_H$ is
evaluated in the QCD perturbative domain where the propagator
virtualities are above the separation scale.

The leading power fall-off of the hard scattering amplitude as
given by dimensional counting rules follows from the nominal
scaling of the hard-scattering amplitude: $T_H \sim 1/Q^{n-4}$,
where $n$ is the total number of fields (quarks, leptons, or gauge
fields) participating in the hard
scattering\cite{Brodsky:1974vy,Matveev:1973ra}. Thus the reaction
is dominated by subprocesses and Fock states involving the minimum
number of interacting fields.  In the case of $2 \to 2$ scattering
amplitudes, this implies $ {d\sigma\over dt}(A B \to C D) ={F_{A B
\to C D}(t/s)/ s^{n-2}}.$ In the case of form factors, the
dominant helicity conserving amplitude obeys $F(t) \sim
(1/t)^{n_H-1}$ where $n_H$ is the minimum number of fields in the
hadron $H$. The full predictions from PQCD modify the nominal
scaling by logarithms from the running coupling and the evolution
of the distribution amplitudes. In some cases, such as large angle
$pp \to p p $ scattering, there can be ``pinch"
contributions\cite{Landshoff:ew} when the scattering can occur
from a sequence of independent near-on shell quark-quark
scattering amplitudes at the same CM angle.  After inclusion of
Sudakov suppression form factors, these contributions also have a
scaling behavior close to that predicted by constituent counting.

The constituent counting rules\cite{Brodsky:1974vy,Matveev:1973ra}
were originally derived in 1973 before the development of QCD, in
anticipation that the underlying theory of hadron physics would be
renormalizable and close to a conformal theory.  The factorizable
structure of hard exclusive amplitudes in terms of a convolution
of valence hadron wavefunctions times a hard scattering quark
scattering amplitude was also proposed.  Upon the discovery of the
asymptotic freedom in QCD, there was a systematical development of
the theory of hard exclusive reactions, including factorization
theorems, counting rules, and evolution equations for the hadronic
distribution
amplitudes\cite{Brodsky:1979qm,Lepage:1979za,Lepage:1979zb,Efremov:1980rn}.
In a remarkable recent development, Polchinski and
Strassler\cite{Polchinski:2001tt} have derived the constituent
counting rules using string duality, mapping features of
gravitational theories in higher dimensions $ (AdS_5)$ to physical
QCD in ordinary 3+1 space-time.

The distribution amplitudes which control leading-twist exclusive
amplitudes at high momentum transfer can be related to the
gauge-invariant Bethe-Salpeter wavefunction at equal light-cone
time $\tau = x^+$.  The logarithmic evolution of the hadron
distribution amplitudes $\phi_H (x_i,Q)$ with respect to the
resolution scale $Q$ can be derived from the
perturbatively-computable tail of the valence light-cone
wavefunction in the high transverse momentum regime. The DGLAP
evolution of quark and gluon distributions can also be derived in
an analogous way by computing the variation of the Fock expansion
with respect to the separation scale. Other key features of the
perturbative QCD analyses are: (a) evolution equations for
distribution amplitudes which incorporate the operator product
expansion, renormalization group invariance, and conformal
symmetry\cite{Lepage:1980fj,Brodsky:1980ny,Muller:1994cn,Ball:1998ff,Braun:1999te};
(b) hadron helicity conservation which follows from the underlying
chiral structure of QCD\cite{Brodsky:1981kj}; (c) color
transparency, which eliminates corrections to hard exclusive
amplitudes from initial and final state interactions at leading
power and reflects the underlying gauge theoretic basis for the
strong interactions\cite{Brodsky:1988xz} and (d) hidden color
degrees of freedom in nuclear wavefunctions, which reflect the
color structure of hadron and nuclear
wavefunctions\cite{Brodsky:1983vf}. There have also been recent
advances eliminating renormalization scale ambiguities in
hard-scattering amplitudes via commensurate scale
relations\cite{Brodsky:1994eh} which connect the couplings
entering exclusive amplitudes to the $\alpha_V$ coupling which
controls the QCD heavy quark potential.

\section{The Pion Form Factor}

The pion spacelike form factor provides an important illustration
of the perturbative QCD formalism. The proof of factorization
begins with the exact Drell-Yan-West
representation\cite{Drell:1970km,West:1970av,Brodsky:1980zm} of
the current in terms of the light-cone Fock wavefunctions (see
Section 7.) The integration over the momenta of the constituents
of each wavefunction can be divided into two domains
$\mathcal{M}^2_n < \Lambda^2 $ and $\mathcal{M}^2_n > \Lambda^2, $
where $\mathcal{M}^2_n$ is the invariant mass of the n-particle
state. $\Lambda$ plays the role of a separation scale.  In
practice, it can be taken to be of order of the momentum transfer.

Consider the contribution of the two-particle Fock state.  The
argument of the final state pion wavefunction is $k_\perp + (1-x)
q_\perp$.  First take $k_\perp$ small.  At high momentum transfer
where
\begin{equation}
\mathcal{M}^2  \sim {(1-x)^2 q^2_\perp \over x(1-x)} = {Q^2
(1-x)\over x} > \Lambda^2,
\end{equation}
one can iterate the equation of motion for the valence light-front
wavefunction using the one gluon exchange kernel.  Including all
of the hard scattering domains, one can organize the result into
the factorized form:
\begin{equation}
F_\pi(Q^2) = \int^1_0 dx \int^1_0 dy \phi_{\pi}(y,\Lambda)
T_H(x,y,Q^2) \phi_{\pi}(x,\Lambda) ,\end{equation} where $T_H$ is
the hard-scattering amplitude $\gamma^* (q \bar q) \to (q \bar q)$
for the production of the valence quarks nearly collinear with
each meson, and $\phi_M(x,\Lambda)$ is the distribution amplitude
for finding the valence $q$ and $\bar q$ with light-cone fractions
of the meson's momentum, integrated over invariant mass up to
$\Lambda.$ The process independent distribution amplitudes contain
the soft physics intrinsic to the nonperturbative structure of the
hadrons.  Note that $T_H$ is non-zero only if ${(1-x)Q^2\over x}
> \Lambda^2 $ and ${(1-y)Q^2\over y} > \Lambda^2 .$ In this
hard-scattering domain, the transverse momenta in the formula for
$T_H$ can be ignored at leading power, so that the structure of
the process has the form of hard scattering on collinear quark and
gluon constituents: $T_H(x,y,Q^2) = {16 \pi C_F
\alpha_s(Q^{*2})\over (1-x) (1-y) Q^2}\left(1 +
\mathcal{O}(\alpha_s)\right)$
and thus\cite{Brodsky:1979qm,Lepage:1979za,Lepage:1979zb,%
Lepage:1980fj,Efremov:1980rn,Chernyak:1977fk,%
Chernyak:1980dk,Farrar:1979aw,Duncan:1980hi}
\begin{equation}
F_{\pi}(Q^2) = {16 \pi C_F \alpha_s(Q^{*2}) \over Q^2}
\int^{\widehat x} _0 dx {\phi_\pi(x,\Lambda)\over (1-x)}
\int^{\widehat y}_0 dy {\phi_\pi(y,\Lambda)\over (1-y)} ,
\end{equation} to leading order in $\alpha_s(Q^{*2})$ and leading
power in $1/Q.$ Here $C_F = 4/3 $ and $Q^*$ can be taken as the
BLM scale\cite{Brodsky:1998dh}. The endpoint regions of
integration $1-x < { \Lambda^2\over Q^2 } = 1-\widehat x $ and
$1-y < { \Lambda^2\over Q^2 }= 1- \widehat y$ are to be explicitly
excluded in the leading-twist formula. However, since the
integrals over $x$ and $y$ are convergent, one can formally extend
the integration range to $0< x < 1$ and $0< y < 1$ with an error
of higher twist. This is only done for convenience -- the actual
domain only encompasses the off-shell regime.  The contribution
from the endpoint regions of integration, $x \sim 1$ and $y \sim
1,$ are power-law and Sudakov suppressed and thus contribute
corrections at higher order in $1/Q$
\cite{Lepage:1979za,Lepage:1979zb,Lepage:1980fj}. The
contributions from non-valence Fock states and corrections from
fixed transverse momentum entering the hard subprocess amplitude
are higher twist, {\em i.e.}, power-law suppressed. Loop
corrections involving hard momenta give next-to-leading-order
(NLO) corrections in $\alpha_s$.

It is sometimes assumed that higher twist terms in the LC wave
function, such as those with $L_z \ne 0$, have flat distributions
at the $ x \to 0,1$ endpoints.  This is difficult to justify since
it would correspond to bound state wavefunctions which fall-off in
transverse momentum but have no fall-off at large $k_z.$ After
evolution to $Q^2 \to \infty$, higher twist distributions can
evolve eventually to constant behavior at $x =0,1;$ however, the
wavefunctions are in practice only being probed at moderate
scales.  In fact, if the higher twist terms are evaluated in the
soft domain, then there is no evolution at all. A recent analysis
by Beneke\cite{Beneke:2002bs} indicates that the $1/Q^4$
contribution to the pion form factor is only logarithmically
enhanced even if the twist-3 term is flat at the endpoints.  It is
also possible that contributions from the twist three  $q \bar q
g$ light-front wavefunctions may well cancel even this
enhancement.

Thus perturbative QCD can unambiguously predict the leading-twist
behavior of exclusive amplitudes.  These contributions only
involve the truncated integration domain of $x$ and $k_\perp$
momenta where the  quark and gluon propagators and couplings are
perturbative; by definition the soft regime is excluded.  The
central question is then whether the PQCD leading-twist prediction
can account for the observed leading power-law fall-off of the
form factors and other exclusive processes.  Assuming the pion
distribution amplitude is close to its asymptotic form,  one can
predict the normalization of exclusive amplitudes such as the
spacelike pion form factor $Q^2 F_\pi(Q^2)$.  Next-to-leading
order predictions are available which incorporate higher order
corrections to the pion distribution amplitude as well as the hard
scattering
amplitude\cite{Muller:1994cn,Muller:1994fv,Melic:1998qr,Szczepaniak:1998sa}.
The natural renormalization scheme for the QCD coupling in hard
exclusive processes is $\alpha_V(Q)$, the effective charge defined
from the scattering of two infinitely-heavy quark test charges.
Assuming $\alpha_V(Q^*) \simeq 0.4$ at the BLM scale $Q^*$, the
QCD LO prediction appears to be smaller by approximately a factor
of 2 compared to the presently available data extracted from pion
electroproduction experiments\cite{Brodsky:1998dh}. However, the
extrapolation from spacelike $t$ to the pion pole in
electroproduction may be unreliable, in the same sense that
lattice gauge theory extrapolations to $m^2_\pi \to 0$ are known
to be nonanalytic.  Thus it is not clear that there is an actual
discrepancy between perturbative QCD and experiment. It would be
interesting to develop predictions for the transition form factor
$F_{q \bar q \to \pi}(t,q^2)$ that is in effect measured in
electroproduction.

Compton scattering is a key test of the perturbative QCD
approach\cite{Kronfeld:1991kp,Guichon:1998xv,Brooks:2000nb}. A
detailed recalculation of the helicity amplitudes and differential
cross section for proton Compton scattering at fixed angle has
been carried out recently by Brooks and Dixon\cite{Brooks:2000nb}
at leading-twist and at leading order in $\alpha_s.$ They use
contour deformations to evaluate the singular integrals in the
light-cone momentum fractions arising from pinch contributions.
The shapes and scaling behavior predicted by perturbative QCD
agree well with the existing data\cite{Shupe:vg}. In order to
reduce uncertainties associated with $\alpha_s$ and the
three-quark wave function normalization, Brooks and Dixon have
normalized the Compton cross section using the proton elastic form
factor. The theoretical predictions for the ratio of Compton
scattering to electron-proton scattering is about an order of
magnitude below existing experimental data. However, this
discrepancy of a factor of 3 in the relative normalization of the
amplitudes could be attributed to the fact that the number of
diagrams contributing to the Compton amplitude at next-to-leading
order ($\alpha_s^3$) is much larger in Compton scattering compared
to the proton form factor.

A debate has continued on whether processes such as the pion and
proton form factors and elastic Compton scattering $\gamma p \to
\gamma p$ might be dominated by higher twist mechanisms until very
large momentum
transfers\cite{Isgur:1989iw,Radyushkin:1998rt,Bolz:1996sw,Diehl:1998kh,Huang:2001ej}.
For example, if one assumes that the light-cone wavefunction of
the pion has the form $\psi_{\rm soft}(x,k_\perp) = A \exp
\left(-b {k_\perp^2\over x(1-x)}\right)$, then the Feynman
endpoint contribution to the overlap integral at small $k_\perp$
and $x \simeq 1$ will dominate the form factor compared to the
hard-scattering contribution until very large $Q^2$.  However,
this form for $\psi_{\rm soft}(x,k_\perp)$ does not fall-off at
all for $k_\perp = 0$ and $k_z \to - \infty$.  A soft QCD
wavefunction would be expected to be exponentially suppressed in
this regime, as in the BHL model $\psi^{\rm soft}_n(x_i, k_{\perp
i}) = A\exp \left(- b\ \sum^n_i [{{\vec k}^2_\perp + m^2\over
x}]_i\right)$ \cite{Lepage:1982gd}. The endpoint contributions are
also suppressed by a QCD Sudakov form factor\cite{Li:1992nu},
reflecting the fact that a near-on-shell quark must radiate if it
absorbs large momentum.  If the endpoint contribution dominates
proton Compton scattering, then both photons will interact on the
same quark line in a local fashion and the amplitude is real, in
strong contrast to the QCD predictions which have a complex phase
structure.  The perturbative QCD
predictions\cite{Kronfeld:1991kp,Guichon:1998xv,Brooks:2000nb} for
the Compton amplitude phase can be tested in virtual Compton
scattering by interference with Bethe-Heitler
processes\cite{Brodsky:1972vv}. Recently the ``handbag" approach
to Compton scattering\cite{Diehl:1998kh,Huang:2001ej} has been
applied to $\bar p p \to \gamma \gamma$ at large
energy\cite{Weiss:2002ec}. In this case, one assumes that the
process occurs via the exchange of a diquark with light-cone
momentum fraction $x \sim 0,$ so that the hard subprocess is $\bar
q q \to \gamma \gamma$ where the quarks annihilate with the full
energy of the baryons and nearly on-shell. The critical question
is whether the proton wavefunction has significant support when
the massive diquark has zero light-front momentum fraction, since
the diquark light-cone kinetic energy and the bound state
wavefunction become far-off shell $k_F^2 \sim - (m^2+k^2_\perp)/x
\to - \infty $ in this domain.

It is interesting to compare the calculation of a meson form
factors in QCD with the calculations of form factors of bound
states in QED. The analog to a soft wavefunction is the
Schr\"odinger-Coulomb solution $\psi_{1s}(\vec k) \propto (1 +
{{\vec p}^2/(\alpha m_{\rm red})^2})^{-2}$, and the full
wavefunction,  which incorporates transversely polarized photon
exchange, differs by a factor $(1 + {\vec p}^2/m^2_{\rm red})$.
Thus the leading-twist dominance of form factors in QED occurs at
relativistic scales $Q^2 > {m^2_{\rm red}}$ \cite{Brodsky:2000dr}.

\section{Perturbative QCD Calculation of Baryon Form Factors}

The baryon form factor at large momentum transfer provides another
important example of the application of perturbative QCD to
exclusive processes.  Away from possible special points in the
$x_i$ integrations (which are suppressed by Sudakov form factors)
baryon form factors can be written to leading order in $1/Q^2$ as
a convolution of a connected hard-scattering amplitude $T_H$
convoluted with the  baryon distribution amplitudes.  The
$Q^2$-evolution of the baryon distribution amplitude can be
derived from the operator product expansion of three quark fields
or from the gluon exchange kernel.  Taking into account the
evolution of the baryon distribution amplitude, the nucleon
magnetic form factors at large $Q^2$, has the
form\cite{Lepage:1980fj,Lepage:1979zb,Brodsky:1981kj}
\begin{equation}
G_M(Q^2)\rightarrow{\alpha^2_s(Q^2)\over Q^4}\sum_{n,m} b_{nm}
\left({\rm log}{Q^2\over \Lambda^2}\right)^{\gamma^B_n+\gamma^B_n}
\left[1+\mathcal{O}\left(\alpha_s(Q^2),{m^2\over
Q^2}\right)\right]\quad .
\end{equation}
where the $\gamma^B_n$ are computable anomalous
dimensions\cite{Peskin:1979mn} of the baryon three-quark wave
function at short distance, and the $b_{mn}$ are determined from
the value of the distribution amplitude $\phi_B(x,Q^2_0)$ at a
given point $Q_0^2$ and the normalization of $T_H$.
Asymptotically, the dominant term has the minimum anomalous
dimension. The contribution from the endpoint regions of
integration, $x \sim 1$ and $y \sim 1,$ at finite $k_\perp$ is
Sudakov suppressed
\cite{Lepage:1979za,Lepage:1979zb,Lepage:1980fj}; however, the
endpoint region may play a significant role in phenomenology.

\begin{figure}[htb]
\begin{center}
\leavevmode
{\epsfxsize=4in\epsfysize=3.5in\epsfbox{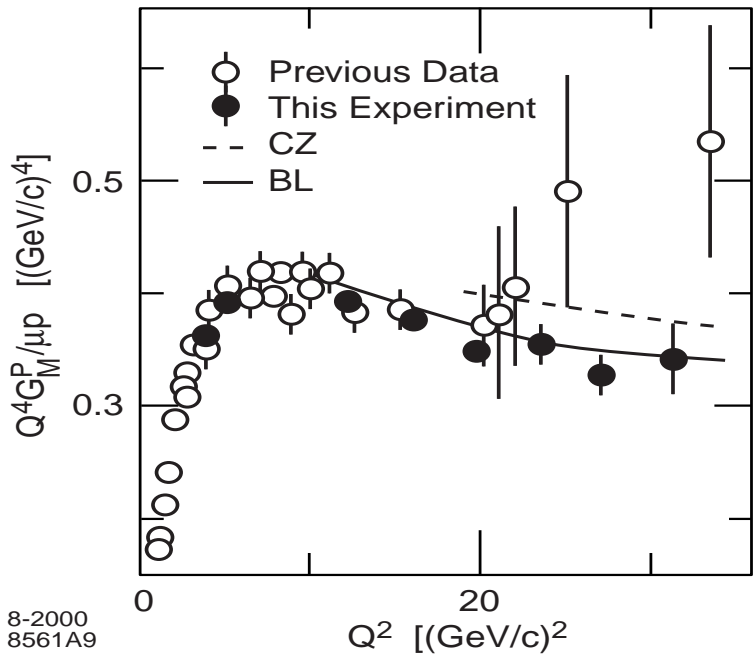}}
\end{center}
\caption[
*]{Predictions for the normalization and sign of the
proton form factor at
high $Q^2$ using perturbative QCD
factorization and QCD sum rule
predictions for the proton
distribution amplitude (From Ji {\em et
al.}\cite{jsl}) The curve
labelled BL has arbitrary normalization and
incorporates the
fall-off of two powers of the running coupling.  The
dotted line
is the QCD sum rule prediction of given by Chernyak
and
Zhitnitsky\cite{Chernyak:1984bm,Chernyak:1989nv}.  The results
are
similar for the model distribution amplitudes of King
and
Sachrajda\cite{ks}, and Gari and
Stefanis\cite{gs}.}
\label{figzrpic}
\end{figure}

The proton
form factor appears to scale at $Q^2 > 5 \ {\rm GeV}^2$
according to the
PQCD predictions.  See Fig. \ref{figzrpic}.
Nucleon form factors are
approximately described
phenomenologically by the well-known dipole form $
G_M(Q^2) \simeq
{1 / (1+Q^2/0.71\,{~\rm GeV}^2)^2}$ which behaves
asymptotically
as $G_M(Q^2) \simeq (1 /Q^4)( 1- 1.42 {~\rm GeV}^2/ Q^2
+
\cdots)\,.$ This suggests that the corrections to leading twist in
the
proton form factor and similar exclusive processes involving
protons become
important in the range $Q^2 < 1.4\ {\rm
GeV}^2$.
\medskip

The shape of the
distribution amplitude controls the normalization
of the leading-twist
prediction for the proton form factor. If one
assumes that the proton
distribution amplitude has the asymptotic
form: $\phi_N = C x_1 x_2 x_3$,
then the convolution with the
leading order form for $T_H$ gives zero! If
one takes a
non-relativistic form peaked at $x_i = 1/3$, the sign is
negative,
requiring a crossing point zero in the form factor at some
finite
$Q^2$.  The broad asymmetric distribution amplitudes advocated
by
Chernyak and Zhitnitsky\cite{Chernyak:1984bm,Chernyak:1989nv}
gives a
more satisfactory result.   If one assumes a constant
value of $\alpha_s =
0.3$, and $f_N=5.3 \times 10^{-3}$GeV$^2$,
the leading order prediction is
below the data by a factor of
$\approx 3.$ However, since the form factor
is proportional to
$\alpha^2_s f^2_N$, one can obtain agreement with
experiment by a
simple renormalization of the parameters.  For example, if
one
uses the central value of Ioffe's determination $f_N=8 \times
10^{-3}
$GeV$^2$, then good agreement is
obtained\cite{Stefanis:1999wy}.  The
normalization of the proton
distribution amplitude is also important for
estimating the proton
decay rate\cite{Brodsky:1983st}. The most recent
lattice
results\cite{Kuramashi:2000hw} suggest a significantly
larger
normalization for the required proton matrix elements, 3 to 5
times
larger than earlier phenomenological estimates.  One can
also use PQCD to
predict ratios of various baryon and isobar form
factors assuming isospin or $SU(3)$-flavor symmetry for the basic
wave function structure.  Results for the neutral weak and charged
weak form factors assuming standard $SU(2)\times U(1)$ symmetry
can also be derived\cite{Brodsky:1981sx}.

A  useful technique for obtaining the solutions to the baryon
evolution equations is to construct completely antisymmetric
representations as a polynomial orthonormal basis for the
distribution amplitude of multi-quark bound states.  In this way
one obtain a distinctive classification of nucleon $(N)$ and Delta
$(\Delta)$ wave functions and the corresponding $Q^2$ dependence
which discriminates $N$ and $\Delta$ form factors.  More recently
Braun and collaborators have shown how one can use conformal
symmetry to classify the eigensolutions of the baryon distribution
amplitude\cite{Braun:1999te}. They identify a new `hidden' quantum
number which distinguishes components in the $\lambda=3/2$
distribution amplitudes with different scale dependence.  They are
able to find analytic solution of the evolution equation for
$\lambda=3/2$ and $\lambda=1/2$ baryons where the  two lowest
anomalous dimensions for the $\lambda=1/2$ operators (one for each
parity) are separated from the rest of the spectrum by a finite
`mass gap'.  These special states can be interpreted as baryons
with scalar diquarks.  Their results may support Carlson's
solution\cite{Carlson:1986mm}  to the puzzle that the proton to
$\Delta$ form factor falls faster\cite{Stoler:1993yk} than other
$p \to N^*$ amplitudes if the $\Delta$ distribution amplitude has
a symmetric $x_1 x_2 x_3$ form.

In a remarkable new development, Pobylitsa {\em et
al.}\cite{Pobylitsa:2001cz} have shown how to compute  transition
form factors linking the proton to nucleon-pion states which have
minimal invariant mass $W$. A new soft pion theorem for high
momentum transfers allows one to compute the three quark
distribution amplitudes for the near threshold pion states from a
chiral rotation. The new soft pion results are in a good agreement
with the SLAC electroproduction data  for $W^2 < 1.4~$GeV$^2$ and
$7 < Q^2 < 30.7~$GeV$^2.$

\section{Hadron Helicity Conservation}

Hadron helicity conservation (HHC) is a QCD selection rule
concerning the behavior of helicity amplitudes at high momentum
transfer, such as fixed CM scattering.  Since the convolution of
$T_H$ with the light-cone wavefunctions projects out states with
$L_z=0$, the leading hadron amplitudes conserve hadron
helicity\cite{Brodsky:1981kj,Chernyak:1999cj}.  Thus the dominant
amplitudes are those in which the sum of hadron helicities in the
initial state equals the sum of hadron helicities in the final
state; other helicity amplitudes are relatively suppressed by an
inverse power in the momentum transfer.

In the case of electron-proton scattering, hadron helicity
conservation states that the proton helicity-conserving form
factor ( which is proportional to $G_M$) dominates over the proton
helicity-flip amplitude  (proportional to $G_E/\sqrt \tau $) at
large momentum transfer.  Here $\tau = Q^2/4M^2, Q^2 = -q^2.$ Thus
HHC predicts  ${G_E(Q^2) / \sqrt \tau G_M(Q^2)} \to 0 $ at large
$Q^2.$  The new data from Jefferson Laboratory\cite{Jones:uu}
which shows a decrease in the ratio ${G_E(Q^2)/  G_M(Q^2)} $ is
not itself in disagreement with the HHC prediction.

\begin{figure}[htbp]
\begin{center}
\leavevmode {\epsfxsize=5in\epsfbox{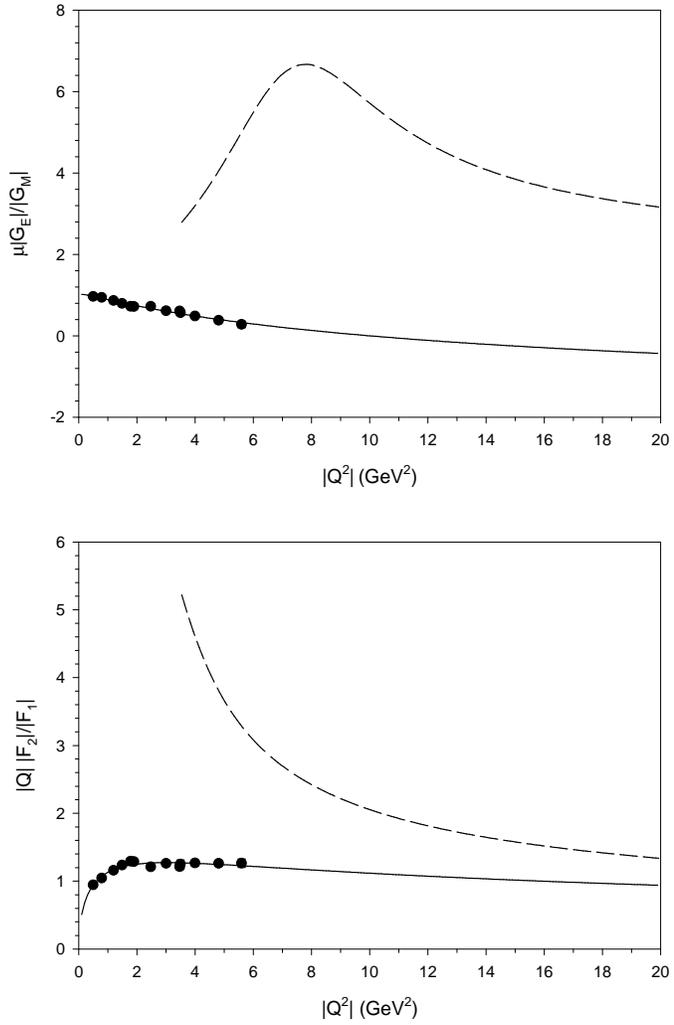}}
\end{center}
\caption[*]{ Perturbative QCD motivated fit\cite{BHH} to the
spacelike form factor ratios and $G_E/G_M$ and $Q F_2/F_1.$ The
fit is described in the text.  The data are from  Jefferson
Laboratory\cite{Jones:uu}.  Predictions for the time-like form
factor ratios are shown as dotted curves. \label{Fig:onelog}}
\end{figure}

The leading-twist QCD motivated form $Q^4 G_M(Q^2) \simeq {{\rm
const} / Q^4 \ln Q^2\Lambda^2}$ provides a good guide to both the
time-like and spacelike proton form factor data at $Q^2
> 5$ GeV$^2$ \cite{Ambrogiani:1999bh}.   However, the Jefferson
Laboratory data\cite{Jones:uu} appears to suggest $Q
F_2(Q^2)/F_1(Q^2) \simeq {\rm const},$  for the ratio of the
proton's  Pauli and Dirac form factors in contrast to the nominal
expectation $Q^2 F_2(Q^2)/F_1(Q^2) \simeq {\rm const}$ expected
(modulo logarithms) from PQCD. It should however be emphasized
that a PQCD-motivated fit is not precluded.  For example, Hiller,
Hwang and I \cite{BHH} have noted that the form
\begin{equation}
{F_2(Q^2)\over F_1(Q^2)} = {\mu_A \over 1 + (Q^2/c) \ln^b(1+
Q^2/a)}
\end{equation}
 with
$\mu_A = 1.79,$ $a = 4 m^2_\pi = 0.073~$GeV$^2,$ $ b = -0.5922,$ $
c = 0.9599~$GeV$^2$ also fits the data well.  The extra
logarithmic factor is not unexpected for higher twist
contributions.  This fit is shown in Fig. \ref{Fig:onelog}.  The
fitted form is consistent with hadron helicity conservation.  The
predictions for the time-like domain using simple crossing of the
above form is shown by the dotted lines.

\vspace{.5cm}

The study of time-like hadronic form factors using $e^+ e^-$
colliding beams can provide very sensitive tests of HHC, since the
virtual photon in $e^+ e^- \to \gamma^* \to h_A \bar h_B$ always
has spin $\pm 1$ along the beam axis at high energies. Angular
momentum conservation implies that the virtual photon can ``decay"
with one of only two possible angular distributions in the center
of momentum frame: $(1+ \cos^2\theta)$ for $\vert\lambda_A -
\lambda_B \vert = 1$ and $\sin^2 \theta$ for $\vert \lambda_A -
\lambda_B \vert = 0$ where $\lambda_A$ and $\lambda_B$ are the
helicities of the outgoing hadrons.  Hadronic helicity
conservation, as required by QCD, greatly restricts the
possibilities.  It implies that $\lambda_A + \lambda_B = 0$.
Consequently, angular momentum conservation requires $\vert
\lambda_A\vert = \vert \lambda_B \vert = l/2$ for baryons, and
$\vert \lambda_A\vert = \vert \lambda_B \vert = 0$ for mesons;
thus the angular distributions for any sets of hadron pairs are
now completely determined at leading twist: $ {d \sigma \over d
\cos\theta}(e^+ e^- = B \bar B) \propto 1 + \cos^2 \theta $ and $
{d \sigma \over d \cos \theta} (e^+ e^- = M \bar M) \propto \sin^2
\theta  .$ Verifying these angular distributions for vector mesons
and other higher spin mesons and baryons would verify the vector
nature of the gluon in QCD and the validity of PQCD applications
to exclusive reactions.

It is usually assumed that a heavy quarkonium state such as the
$J/\psi$ always decays to light hadrons via the annihilation of
its heavy quark constituents to gluons.  However, as Karliner and
I \cite{Brodsky:1997fj} have shown, the transition $J/\psi \to
\rho \pi$ can also occur by the rearrangement of the $c \bar c$
from the $J/\psi$ into the $\ket{ q \bar q c \bar c}$ intrinsic
charm Fock state of the $\rho$ or $\pi$.  On the other hand, the
overlap rearrangement integral in the decay $\psi^\prime \to \rho
\pi$ will be suppressed since the intrinsic charm Fock state
radial wavefunction of the light hadrons will evidently not have
nodes in its radial wavefunction.  This observation provides a
natural explanation of the long-standing puzzle why the $J/\psi$
decays prominently to two-body pseudoscalar-vector final states in
conflict with HHC, whereas the $\psi^\prime$ does not. If the
intrinsic charm explanation is correct, then this mechanism will
complicate the analysis of virtually all heavy hadron decays such
as $J/\psi \to p \bar p.$ In addition, the existence of intrinsic
charm Fock states, even at a few percent level, provides new,
competitive decay mechanisms for $B$ decays which are nominally
CKM-suppressed\cite{Brodsky:2001yt}. For example, the weak decays
of the B-meson to two-body exclusive states consisting of strange
plus light hadrons, such as $B\to\pi K,$ are expected to be
dominated by penguin contributions since the tree-level $b\to s
u\bar u$ decay is CKM suppressed. However, higher Fock states in
the B wave function containing charm quark pairs can mediate the
decay via a CKM-favored $b\to s c\bar c$ tree-level transition.
The presence of intrinsic charm in the $b$ meson can be checked by
the observation of final states containing three charmed quarks,
such as $B \to J/\psi D \pi$ \cite{Chang:2001yf}.

\section{Other Applications}

There are a large number of measured exclusive reactions in which
the
empirical power law fall-off predicted by dimensional counting
and PQCD
appears to be accurate over a large range of
momentum
transfer.
\vspace{.5cm}
\begin{figure}[htb]
\begin{center}
\leavevmode {\epsfxsize=4.5in\epsfysize=4in\epsfbox{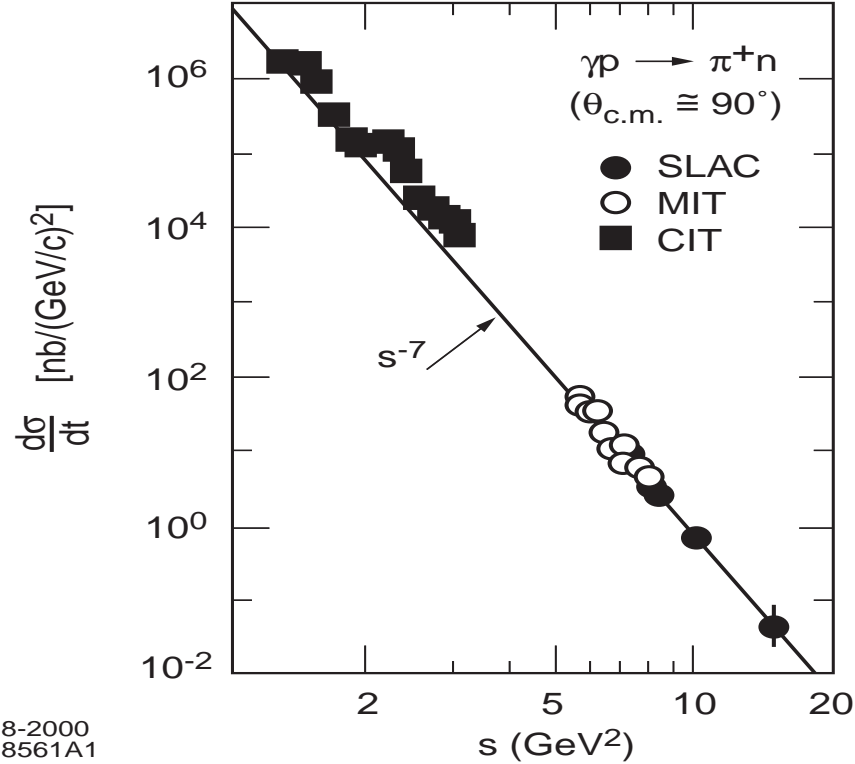}}
\end{center}
\caption[
*]{Comparison of photoproduction data with the
dimensional counting
power-law prediction.  The data are
summarized in Anderson {\em et
al.}\cite{Arr}} \label{figzkpic}
\end{figure}
The approach to scaling of
$s^7 d\sigma/dt(\gamma p \to \pi^+ n)$
shown in Fig. \ref{figzkpic} appears
to indicate that
leading-twist PQCD is applicable at momentum transfers
exceeding a
few GeV.  If anything, the scaling appears to work too
well,
considering that one expects logarithmic deviations from the
running
of the QCD coupling and logarithmic evolution of the
hadron distribution
amplitudes.   The deviations from scaling at
lower energies\cite{Besch:sx}
are interesting and can be
attributed to $s$-channel resonances or perhaps
heavy quark
threshold effects, merging into the fixed-angle scaling in
a
similar way as one observes the approach to leading-twist
Bjorken-scaling
behavior in deep inelastic scattering via
quark-hadron
duality\cite{Melnitchouk:2001zy}.  The absence of
significant corrections
to leading-twist scaling suggests that the
running coupling is effectively
frozen at the kinematics relevant
to the data.   If higher-twist soft
processes are conspiring to
mimic leading-twist scaling $s^7
d\sigma/dt(\gamma p \to \pi^+n)$,
then we would have the strange situation
of seeing two separate
kinematic domains of $s^7$ scaling of the
photoproduction cross
section. It has been
argued\cite{Radyushkin:1998rt,Diehl:2001fv}
that the Compton amplitude is
dominated by soft end-point
contributions of the proton wavefunctions where
the two photons
both interact on a quark line carrying nearly all of the
proton's
momentum.  However, a corresponding soft endpoint explanation
of
the observed $s^7 d\sigma/dt(\gamma p \to \pi^+ n)$ scaling of the
pion
photoproduction data is not apparent;  there is no endpoint
contribution
which could explain the success of dimensional
counting in large-angle pion
photoproduction apparent in Fig.
\ref{figzkpic}.

Exclusive two-photon
processes where two photons annihilate into
hadron pairs $\gamma \gamma \to
H \bar H$ at high transverse
momentum  provide highly valuable probes of
coherent effects in
quantum chromodynamics.  For example, in the case of
exclusive
final states at high momentum transfer and fixed $\theta_{cm}$
such as $\gamma \gamma \rightarrow p \bar p $ or meson pairs,
photon-photon collisions provide a time-like microscope for
testing fundamental scaling laws of PQCD and for measuring
distribution amplitudes. Counting rules predict asymptotic
fall-off $s^4 d\sigma/dt \sim f(t/s)$ for meson pairs and  $s^6
d\sigma/dt \sim f(t/s)$ for baryon pairs.  Hadron-helicity
conservation predicts dominance of final states with $\lambda_H +
\lambda_{\bar H} = 0.$  The angular dependence reflects the
distribution amplitudes.  One can also study $\gamma^* \gamma \to
$ hadron pairs in $ e^\pm e^-$ collisions as a function of photon
virtuality, the time-like analog of deeply virtual Compton
scattering which is sensitive to the two hadron distribution
amplitude.  One can also study the interference of the time-like
Compton amplitude with the bremsstrahlung amplitude $e^\pm e \to B
B e^\pm e^-.$ where a time-like photon produces the pair.  The
$e^\pm$ asymmetry measures the relative phase of the time-like
hadron form factor with that of the virtual Compton amplitude.

\begin{figure}[htbp]
\begin{center}
\leavevmode {\epsfxsize=4in\epsfbox{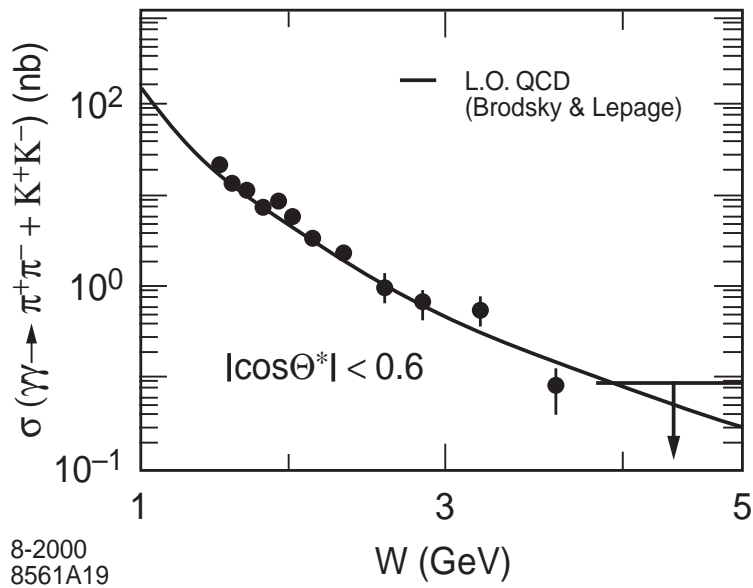}}
\end{center}
\caption[*]{Comparison of the sum of $\gamma \gamma \rightarrow
\pi^+ \pi^-$ and $\gamma \gamma \rightarrow K^+ K^-$ meson pair
production cross sections with the perturbative QCD
prediction\cite{BL} normalized to the timelike pion
form factor. The data are from the CLEO collaboration\cite{Paar}.}
\label{Fig:CLEO}
\end{figure}

The PQCD predictions for the two-photon production of charged
pions and kaons is insensitive to the shape of the meson
distribution amplitudes.  In fact, the ratio of the $\gamma \gamma
\to \pi^+ \pi^-$ and  $e^+ e^- \to \mu^+ \mu^-$ amplitudes at
large $s$ and fixed $\theta_{CM}$ can be predicted since the ratio
is nearly insensitive to the running coupling and the shape of the
pion distribution amplitude:
\begin{equation}
{{d\sigma \over dt }(\gamma \gamma \to \pi^+ \pi^-) \over {d\sigma
\over dt }(\gamma \gamma \to \mu^+ \mu^-)} \sim {4 \vert F_\pi(s)
\vert^2 \over 1 - \cos^2 \theta_{\rm c.m.} } .\end{equation} The
comparison of the PQCD prediction for the sum of $\pi^+ \pi^-$
plus $K^+ K^-$ channels with recent CLEO data\cite{Paar} is shown
in Fig. \ref{Fig:CLEO}.  Results for separate pion and kaon
channels have been given by the TPC/$2\gamma$
collaboration\cite{Boyer}.
\vspace{.5cm}
The angular distribution of meson pairs is also predicted by PQCD
at large momentum transfer. The CLEO data for charged pion and
kaon pairs show a clear transition to the angular distribution
predicted by PQCD for $W = \sqrt s_{\gamma \gamma} > 2$ GeV.
Similarly in $\gamma \gamma \to p \bar p$ one can see a dramatic
change in the fixed angle distribution as one enters the hard
scattering domain. It is clearly important to measure the
two-photon production of neutral pions and $\rho^+ \rho^-$ cross
sections in view of their strong sensitivity to the shape of meson
distribution amplitudes. Furthermore, the ratio of $\pi^+ \pi^-$
to $\pi^0 \pi^0$ cross sections is highly sensitive to the
production dynamics. The ratio ${\sigma(\gamma \gamma \to \pi^0
\pi^0)\over \sigma(\gamma \gamma \to \pi^+ \pi^-)}$ at fixed
angles which is very small in PQCD, and of order $1$ in soft
handbag models.

An interesting contribution to $K^+ p \to K^+ p$ scattering comes
from the exchange of the common $u$ quark.  The  quark interchange
amplitude for $A + B \to  C + D$ can be written as a convolution
of the four light-cone wavefunctions multiplied by a factor
$\Delta^- = P^-_A + P^-_B - \sum_i k^-_i,$  the inverse of the
central propagator\cite{Gunion:1973ex}. The interchange amplitude
is consistent with constituent counting rule scaling, and often
provides a phenomenologically accurate representation of the
$\theta_{c.m.}$ angular distribution at large momentum transfer.
For example, the angular distribution of processes such as $K^+ p
\to K^+ p$ appear to follow the predictions based  on hard
scattering diagrams based on quark interchange, \eg, $T_H((u_1
\bar s) (u_2 u_3 d) \to (u_2 \bar s) (u_1 u_3 d)$
\cite{Gunion:1973ex}.  This mechanism also provides constraints on
Regge intercepts $\alpha_R(t)$ for  meson exchange trajectories at
large momentum transfer\cite{Blankenbecler:1973kt}.  An extensive
review of this phenomenology is given in the review by Sivers {\em
et al.}\cite{Sivers:1976dg}

\begin{figure}[htb]
\begin{center}
\leavevmode {\epsfxsize=4in\epsfbox{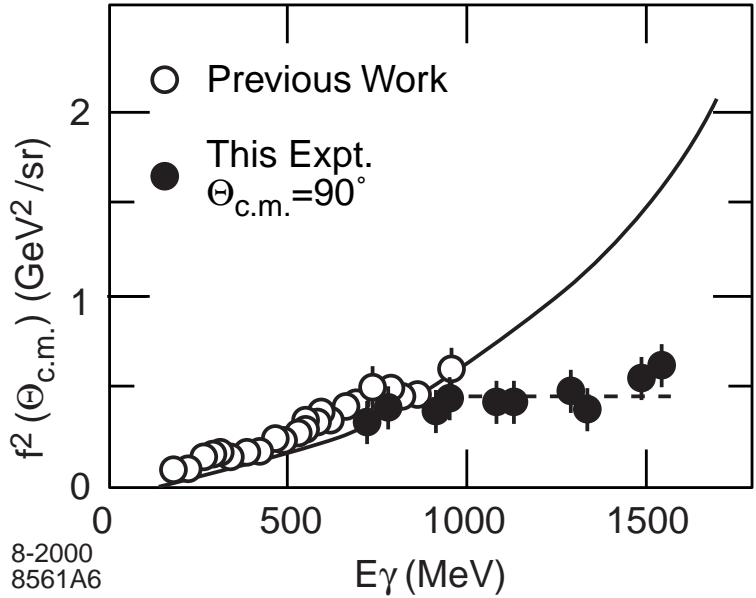}}
\end{center}
\caption[*]{ Comparison of deuteron photodisintegration data with
the scaling prediction which requires $f^2(\theta_{cm})$ to be at
most logarithmically dependent on energy at large momentum
transfer.  The data in are from Belz {\em et
al.}\cite{Belz:1995ge} The solid curve is a nuclear physics
prediction\cite{Lee:1988pi}.} \label{figyqpic}
\end{figure}

One of the most interesting areas of exclusive processes is to
amplitudes where the nuclear wavefunction has to absorb large
momentum transfer. For example, the helicity-conserving deuteron
form factor is predicted to scale as $F_d(Q^2) \propto (Q^2)^{-5}$
reflecting the minimal six quark component of nuclear
wavefunction. The deuteron form factor at high $Q^2$ is sensitive
to wavefunction configurations where all six quarks overlap within
an impact separation $b_{\perp i} < \mathcal{O} (1/Q).$  The
leading power-law fall off predicted by QCD is $F_d(Q^2) =
f(\alpha_s(Q^2))/(Q^2)^5$, where,
asymptotically\cite{Brodsky:1976rz,Brodsky:1983vf},
$f(\alpha_s(Q^2)) \propto \alpha_s(Q^2)^{5+2\gamma}$.  In general,
the six-quark wavefunction of a deuteron is a mixture of five
different color-singlet states.  The dominant color configuration
at large distances corresponds to the usual proton-neutron bound
state. However, at small impact space separation, all five Fock
color-singlet components eventually acquire equal weight, \ie, the
deuteron wavefunction evolves to 80\%\ ``hidden
color''\cite{Brodsky:1983vf}. The relatively large normalization
of the deuteron form factor observed at large $Q^2$ hints at
sizable hidden-color contributions\cite{Farrar:1991qi}. Hidden
color components can also play a predominant role in the reaction
$\gamma d \to J/\psi p n$ at threshold if it is dominated by the
multi-fusion process $\gamma g g \to J/\psi$.  In the case of
nuclear structure functions beyond the single nucleon kinematic
limit, $1 < x_{bj} < A$, the nuclear light-cone momentum must be
transferred to a single quark, requiring quark-quark correlations
between quarks of different nucleons in a compact, far-off-shell
regime.  This physics is also sensitive to the part of the nuclear
wavefunction which contains hidden-color components in distinction
from a convolution of separate color-singlet nucleon
wavefunctions. One also sees the onset of the predicted
perturbative QCD scaling behavior for exclusive nuclear amplitudes
such as deuteron photodisintegration ($n = 1+ 6 + 3 + 3 = 13$)
$s^{11}{ d\sigma\over dt}(\gamma d \to p n) \sim $ constant at
fixed CM angle. The measured deuteron form factor and the deuteron
photodisintegration cross section appear to follow the
leading-twist QCD predictions at large momentum transfers in the
few GeV region\cite{Holt:1990ze,Bochna:1998ca}. To first
approximation, the proton and neutron share the deuteron's
momentum equally.  Since the deuteron form factor contains the
probability amplitudes for the proton and neutron to scatter from
$p/2$ to $p/2+q/2$; it is natural to define the reduced deuteron
form factor\cite{Brodsky:1976rz,Brodsky:1983vf}
\begin{equation} f_d(Q^2) \equiv {F_d(Q^2)\over
F_{1N} \left(Q^2\over 4\right)\, F_{1N}\,\left(Q^2\over
4\right)}.\end{equation} The effect of nucleon compositeness is
removed from the reduced form factor.  QCD then predicts the
scaling
\begin{equation} f_d(Q^2) \sim {1\over Q^2} ; \end{equation}
\ie\ the same scaling law as a meson form factor. This scaling is
consistent with experiment for $Q \gsim$ 1 GeV.  In the case of
deuteron photodisintegration $\gamma d \to p n$ the amplitude
requires the scattering of each nucleon at $t_N = t_d/4$.  The
perturbative QCD scaling is\cite{Brodsky:1983kb}
\begin{equation}
{d\sigma\over d\Omega_{c.m.}} (\gamma d \to n p) = {1\over \sqrt
{s (s-M_d^2)}}
 {F^2_n(t_d/4)  F^2_p(t_d/4) f^2_{red}(\theta_{c.m})\over p^2_\perp} .
\end{equation}
The predicted scaling of the reduced photodisintegration amplitude
$f_{red}(\theta_{c.m.})  \simeq $ const  is also consistent with
experiment\cite{Brodsky:1983kb,Holt:1990ze,Bochna:1998ca}.  See
Fig. \ref{figyqpic}.

The postulate that the QCD coupling has an infrared fixed-point
provides an understanding of the applicability of conformal
scaling and constituent counting rules to physical QCD
processes\cite{Brodsky:1974vy,Matveev:1973ra}. The general success
of dimensional counting rules implies that the effective coupling
$\alpha_V(Q^*)$ controlling the gluon exchange propagators in
$T_H$ are frozen in the infrared, since the effective momentum
transfers $Q^*$ exchanged by the gluons are often a small fraction
of the overall momentum transfer\cite{Brodsky:1998dh}. In this
case, the pinch contributions are suppressed by a factor
decreasing faster than a fixed power\cite{Brodsky:1974vy}. The
effective coupling $\alpha_\tau(s)$ extracted from $\tau$ decays
displays a flat behavior at low mass scales\cite{Menke}.

The field of analyzable exclusive processes has been expanded to a
wide range of QCD processes, such as electroweak decay amplitudes,
highly virtual diffractive processes such as $\gamma^* p \to \rho
p$ \cite{Brodsky:1994kf,Collins:1997sr}, and semi-exclusive
processes such as $\gamma^* p \to \pi^+ X$ \cite{Brodsky:1998sr}
where the $\pi^+$ is produced in isolation at large $p_T$.  An
important new application is the recent analysis of hard exclusive
$B$ decays by Beneke {\em et al.}\cite{Beneke:2000ry} and Keum
{\em et al.}\cite {Keum:2000wi}  Deeply virtual Compton amplitude
$\gamma^* p \to \gamma p$ has emerged as one of the most important
exclusive QCD
reactions\cite{Ji:1997nm,Radyushkin:1997ki,Diehl:1999tr,Diehl:1999kh}.
The process factorizes into a hard amplitude representing Compton
scattering on the quark times skewed parton distributions.  The
resulting skewed parton form factors can be represented as
diagonal and off-diagonal convolutions of light-cone
wavefunctions, as in semileptonic $B$ decay\cite{Brodsky:2000xy}.
New sum rules can be constructed which correspond to gravitons
coupling to the quarks of the proton\cite{Ji:1997nm}.  It is
possible that the handbag approximation to DVCS may be modified by
corrections to the quark propagator similar to those which appear
in the final state interaction corrections to deep inelastic
scattering\cite{Brodsky:2002ue,Brodsky:2000ii}.  In particular,
one can expect that the propagator corrections will give
single-spin asymmetries correlating the spin of the proton with
the normal to the production plane in DVCS \cite{Brodsky:2002cx}.

The hard diffraction of vector mesons $\gamma^* p \to V^0 p$ at
high $Q^2$ and high energies for longitudinally polarized vector
mesons factorizes into a skewed parton distribution times the hard
scale $\gamma^* g \to g V^0$ amplitude, where the physics of the
vector meson is contained in its distribution
amplitude\cite{Brodsky:1994kf,Muller:1994cn,Collins:1997fb}. The
data appears consistent with the $s,t$ and $Q^2$ dependence
predicted by  theory.  Ratios of these processes for different
mesons are sensitive to the ratio  of $1/x$ moments of the $V^0$
distribution amplitudes.

The two-photon annihilation process $\gamma^* \gamma \to $
hadrons, which is measurable in single-tagged $e^+ e^- \to e^+ e^-
{\rm hadrons}$ events provides a semi-local probe of even charge
conjugation $C=^+$ hadron systems $\pi^0, \eta^0, \eta^\prime,
\eta_c, \pi^+ \pi^-$, etc.   The   $\gamma^* \gamma \to \pi^+
\pi^-$ hadron pair process is related to virtual Compton
scattering on a pion target by crossing.  Hadron pair production
is of particular interest since the leading-twist amplitude is
sensitive to the $1/x - 1/(1-x)$ moment of the two-pion
distribution amplitude coupled to two valence quarks%
\cite{Muller:1994cn,Diehl:2000uv}.   This type of measurement can
also constrain the parameters of the effective chiral theory,
which is needed for example to constrain the hadronic
light-by-light contribution to the muon magnetic
moment\cite{Ramsey-Musolf:2002cy}.

One can also study hard ``semi-exclusive''
processes\cite{Brodsky:1998sr} of the form $A+B \to C + Y$ which
are characterized by a large momentum transfer between the
particles $A$ and $C$ and a large rapidity gap between the final
state particle $C$ and the inclusive system $Y$. Such reactions
are in effect generalizations of deep inelastic lepton scattering,
providing novel currents which probe specific quark distributions
of the target $B$ at fixed momentum fraction and novel
spin-dependent parton distributions.

\section{Exact Formulae for Exclusive Processes}

The natural formalism for describing the hadronic wavefunctions
which enter exclusive and diffractive amplitudes is the light-cone
Fock representation obtained by quantizing the theory at fixed
``light-cone" time $\tau = t+z/c$~\cite{Brodsky:1998de}. For
example, the proton state has the Fock expansion
\begin{eqnarray}
\ket p &=& \sum_n \VEV{n\,|\,p}\, \ket n \nonumber \\
&=& \psi^{(\Lambda)}_{3q/p} (x_i,\vec k_{\perp i},\lambda_i)\,
\ket{uud} \\[1ex]
&&+ \psi^{(\Lambda)}_{3qg/p}(x_i,\vec k_{\perp i},\lambda_i)\,
\ket{uudg} + \cdots \nonumber \label{eq:b}
\end{eqnarray}
representing the expansion of the exact QCD eigenstate on a
non-interacting quark and gluon basis.  The probability amplitude
for each such $n$-particle state of on-mass shell quarks and
gluons in a hadron is given by a light-cone Fock state
wavefunction $\psi_{n/H}(x_i,\vec k_{\perp i},\lambda_i)$, where
the constituents have longitudinal light-cone momentum fractions $
x_i ={k^+_i}/{p^+} = (k^0_i+k^z_i)/(p^0+p^z)\ , \sum^n_{i=1} x_i=
1 $, relative transverse momentum $\vec k_{\perp i} \ ,
\sum^n_{i=1}\vec k_{\perp i} = \vec 0_\perp$, and helicities
$\lambda_i.$ The effective lifetime of each $n-$ parton
configuration in the laboratory frame is $2 P_{lab}\over \M^2_n-
M^2_p $ where ${\M}^2_n = \sum^n_{i=1}(k^2_{\perp i} + m^2_i)/x_i
< \Lambda^2 $ is the off-shell invariant mass and $\Lambda$ is a
global ultraviolet regulator.  A crucial feature of the light-cone
formalism is the fact that the form of the
$\psi^{(\Lambda)}_{n/H}(x_i, \vec k_{\perp i},\lambda_i)$ is
invariant under longitudinal boosts; \ie,\ the light-cone
wavefunctions expressed in the relative coordinates $x_i$ and
$k_{\perp i}$ are independent of the total momentum $P^+$, $\vec
P_\perp$ of the hadron. Angular momentum conservation also has a
precise meaning in the light-front representation: for each
$n-$particle Fock state,
\begin{equation}
J^z = \sum^n_{i =1} S^z_i + \sum^{n-1}_i L^z_i
\end{equation}
since there are only $n-1$ relative orbital angular
momenta\cite{Brodsky:2000ii}.

Matrix elements of space-like local operators for the coupling of
photons, gravitons and the deep inelastic structure functions can
all be expressed as overlaps of light-cone wavefunctions with the
same number of Fock constituents.  This is possible since one can
choose the special frame $q^+ = 0$ \cite{Drell:1970km,West:1970av}
for space-like momentum transfer and take matrix elements of
``plus" components of currents such as $J^+$ and $T^{++}$.  Since
the physical vacuum in light-cone quantization coincides with the
perturbative vacuum, no contributions to matrix elements from
vacuum fluctuations occur\cite{Brodsky:1998de}. In the light-cone
formalism one can identify the Dirac and Pauli form factors from
the light-cone spin-conserving and spin-flip vector current matrix
elements of the $J^+$ current\cite{Brodsky:1980zm}: $
\VEV{P+q,\uparrow\left|\frac{J^+(0)}{2P^+} \right|P,\uparrow}
=F_1(q^2) \ , $ $
\VEV{P+q,\uparrow\left|\frac{J^+(0)}{2P^+}\right|P,\downarrow}
=-{(q^1-{\mathrm i} q^2)\over 2M}{F_2(q^2)}\ . $ More explicitly,
the Pauli form factor  can  be  calculated from the expression
\begin{equation}
-(q^1-{\mathrm i} q^2){F_2(q^2)\over 2M} = \sum_a  \int {{\mathrm
d}^2 {\vec k}_{\perp} {\mathrm d} x \over 16 \pi^3} \sum_j e_j \
\psi^{\uparrow *}_{a}(x_i,{\vec k}^\prime_{\perp i},\lambda_i) \,
\psi^\downarrow_{a} (x_i, {\vec k}_{\perp i},\lambda_i) {}\ ,
\label{LCmu}
\end{equation}
where  the summation is over all contributing Fock states $a$ and
struck constituent charges $e_j$.  The arguments of the
final-state light-cone  wavefunction
are\cite{Drell:1970km,West:1970av} $ {\vec k}'_{\perp i}={\vec
k}_{\perp i}+(1-x_i){\vec q}_{\perp} $ for the struck constituent
and $ {\vec k}'_{\perp i}={\vec k}_{\perp i}-x_i{\vec q}_{\perp} $
for each spectator.  The Pauli form factor couples Fock states
differing by one unit of orbital angular momentum, since the
initial and final states have opposite total spin $J_z$ and the
constituent spins are unchanged in the overlap formula. In effect,
one is measuring the spin-orbit $\vec S \cdot \vec L$ in the
light-front formalism. Thus the $F_2$ form factor and the
anomalous moment directly measure relative angular momentum in the
proton. This is also true for the $E$ generalized parton
distribution determined in DVCS and the single-spin asymmetries
measured in DIS. In these cases, the quark contributions are
weighted by the square of the quark charges.

In the ultra-relativistic limit where the radius of the system is
small compared to its Compton scale $1/M$, the anomalous magnetic
moment must vanish\cite{Bro94}.  The light-cone formalism is
consistent with this theorem. The anomalous moment coupling $B(0)$
to a graviton vanishes for any composite system. This remarkable
result, first derived by Okun and
Kobzarev\cite{Okun,Ji:1996ek,Ji:1997pf,Teryaev:1999su}, follows
directly from the Lorentz boost properties of the light-cone Fock
representation\cite{Brodsky:2000ii}.

The overlap formula for the form factors is invariant under $\vec
q_\perp \to - \vec q _\perp.$   Thus at large momentum transfer
one obtains an expansion of form factors in powers of $1/q^2$
modulo logarithms.  This also can be seen from a twist expansion
of the operator product expansion\cite{Braun:2001tj}.

Exclusive semi-leptonic $B$-decay amplitudes involving time-like
currents such as $B\rightarrow A \ell \bar{\nu}$ can also be
evaluated exactly in the light-cone
framework\cite{Brodsky:1999hn,Ji:1999gt}. In this case, the $q^+ =
0$ frame cannot be used, and the time-like decay matrix elements
require the computation of both the diagonal matrix element $n
\rightarrow n$ where parton number is conserved and the
off-diagonal $n+1\rightarrow n-1$ convolution such that the
current operator annihilates a $q{\bar{q'}}$ pair in the initial
$B$ wavefunction.  See Fig. 2(h). A similar result holds for the
light-cone wavefunction representation of the deeply virtual
Compton amplitude\cite{Brodsky:2000xy}.  This feature will carry
over to exclusive hadronic $B$-decays, such as $B^0 \rightarrow
\pi^-D^+$. In this case the pion can be produced from the
coalescence of a $d\bar u$ pair emerging from the initial higher
particle number Fock wavefunction of the $B$.  The $D$ meson is
then formed from the remaining quarks after the internal exchange
of a $W$ boson.

In principle, a precise evaluation of the hadronic matrix elements
needed for $B$-decays and other exclusive electroweak decay
amplitudes requires knowledge of all of the light-cone Fock
wavefunctions of the initial and final state hadrons.  In the case
of model gauge theories such as QCD(1+1) \cite{Horn} or collinear
QCD \cite{AD} in one-space and one-time dimensions, the complete
evaluation of the light-cone wavefunction is possible for each
baryon or meson bound-state using the DLCQ method\cite{DLCQ,AD}.
It would be interesting to use such solutions as a model for
physical $B$-decays.

There are now real prospects of computing the hadron wavefunctions
and distribution amplitudes from first principles in QCD as
exemplified by the computation\cite{Dalley:2000dh} of the pion
distribution amplitude using a combination of DLCQ and the
transverse lattice methods and recent results from traditional
lattice gauge theory\cite{DelDebbio:2000mq}.  Instanton models
predict a pion distribution amplitude close to the asymptotic
form\cite{Petrov:1999kg}. A new result for the proton distribution
amplitude treating nucleons as chiral solitons has recently been
derived by  Diakonov and Petrov\cite{Diakonov:2000pa}.
Dyson-Schwinger models\cite{Hecht:2000xa} of hadronic
Bethe-Salpeter wavefunctions can also be used to predict
light-cone wavefunctions and hadron distribution amplitudes by
integrating over the relative $k^-$ momentum.

\section{Color Transparency}

Each hadron entering or emitted from a hard exclusive reaction
initially emerges with high momentum and small transverse size
$b_\perp =\mathcal{O}(1/\widetilde Q)$.  A fundamental feature of
gauge theory is that soft gluons decouple from the small
color-dipole moment of the compact fast-moving color-singlet
wavefunction configurations of the incident and final-state
hadrons. The transversely compact color-singlet configurations can
effectively persist over a distance of order $\ell_{\rm Ioffe} =
\mathcal{O} (E_{\rm lab}/Q^2)$, the Ioffe coherence length.  Thus
if we study hard quasi-elastic processes in a nuclear target such
as $e A \to e' p' (A-1)$ or $p A \to p' (A-1)$, the outgoing and
ingoing hadrons will have minimal absorption in a nucleus.  The
diminished absorption of hadrons produced in hard exclusive
reactions implies additivity of the nuclear cross section in
nucleon number $A$ and is the theoretical basis for the ``color
transparency" of hard quasi-elastic reactions%
\cite{Brodsky:1988xz,Frankfurt:1988nt,Jain:1996dd,Griff}.  In
contrast, in conventional Glauber scattering, one predicts strong,
nearly energy-independent initial and final state attenuation.
Similarly, in hard diffractive processes such as $\gamma^*(Q^2) p
\to \rho p$ \cite{Brodsky:1994kf} only the small transverse
configurations $b_\perp \sim 1/Q$ of the longitudinally polarized
vector meson distribution amplitude is involved. Its hadronic
interactions as it exits the nucleus will be minimal, and thus the
$\gamma^*(Q^2) N \to \rho N$ reaction can occur coherently
throughout a
nuclear target in reactions without absorption or
shadowing.  Evidence for
color transparency in such reactions has
been reported by Fermilab
experiment E665\cite{Adams:1994bw}.

The most convincing demonstration of
color transparency has been
reported by the E791 group at
FermiLab\cite{Aitala:2000hc} in
measurements of diffractive dissociation of
a high energy pions to
high transverse momentum dijets; $\pi A \to jet\
jet\  A$; the
forward diffractive amplitude is observed to grow in
proportion to
the total number of nucleons in the nucleus, in strong
contrast to
standard Glauber theory which predicts that only the front
surface
of the nucleus should be effective.

There is  also evidence for
the onset of color transparency in
large angle quasi-elastic $p p$
scattering in
nuclear
targets\cite{Carroll:1988rp,Mardor:1998fz,Leksanov:2001ui}, in
the
regime $6< s < 25$ GeV$^2$, indicating that small
wavefunction
configurations are indeed controlling this exclusive reaction
at
moderate momentum transfers.  However at $p_{\rm lab} \simeq 12$
GeV,
$E_{cm} \simeq 5$ GeV, color transparency dramatically fails.
It is noteworthy that in the same energy range, the normal-normal
spin asymmetry $A_{NN}$ in elastic $pp \to pp$ scattering at
$\theta_{cm} = 90^0$ increases dramatically to $A_{NN} \simeq 0.6$
 -- it is about four times more probable that the protons scatter
with helicity normal to the scattering plane than
anti-normal\cite{Court:1986dh}.

The unusual spin and color transparency effects seen in elastic
proton-proton scattering at $E_{CM} \sim 5$ GeV and large angles
could be related to the charm threshold and the effects of a
$\ket{ uud uud c \bar c }$ resonance which would appear as in the
$J=L=S=1$ $p p $ partial
wave\cite{Brodsky:1988xw,deTeramond:1998ny}. The intermediate
state $\vert u u d u u d c \bar c \rangle$ has odd intrinsic
parity and couples to the $J=S=1$ initial state, thus strongly
enhancing scattering when the incident projectile and target
protons have their spins parallel and normal to the scattering
plane.  A similar enhancement of $A_{NN}$ is observed at the
strangeness threshold.  The physical protons coupling at the charm
threshold will have normal Glauber interactions, thus explaining
the anomalous change in color transparency observed at the same
energy in quasi-elastic $ p p$ scattering.  A crucial test of the
charm hypothesis is the observation of open charm production near
threshold with a cross section of order of $1 \mu$b
\cite{Brodsky:1988xw,deTeramond:1998ny}. A similar cross section
is expected for the second threshold for open charm production
from $p \bar p \to {\rm charm} ~ p \bar p.$ An alternative
explanation of the color transparency and spin anomalies in $pp$
elastic scattering has been postulated by Ralston, Jain, and
Pire\cite{Ralston:1986zn,Jain:1996dd}. The oscillatory effects in
the large-angle $pp \to pp$ cross section and spin structure are
postulated to be due to the interference of Landshoff pinch and
perturbative QCD amplitudes.  In the case of quasi-elastic
reactions, the nuclear medium absorbs and filters out the
non-compact pinch contributions, leaving the additive hard
contributions unabsorbed.  It is clearly important that these two
alternative explanations be checked by experiment.

In general, one can expect strong effects whenever heavy quarks
are produced at low relative velocity with respect to each other
or the other quarks in the reaction since the QCD van der Waals
interactions become maximal in this domain. The opening of the
strangeness and charm threshold in intermediate states can become
most apparent in large angle reactions such as $pp$ scattering and
pion photoproduction since the competing perturbative QCD
amplitudes are power-suppressed. Charm and bottom production near
threshold such as $J/\psi$ photoproduction is also sensitive to
the multi-quark, gluonic, and hidden-color correlations of
hadronic and nuclear wavefunctions in QCD since all of the
target's constituents must act coherently within the small
interaction volume of the heavy quark production
subprocess\cite{Brodsky:2000zc}. Although such multi-parton
subprocess cross sections are suppressed by powers of $1/m^2_Q$,
they have less phase-space suppression and can dominate the
contributions of the leading-twist single-gluon subprocesses in
the threshold regime.

\section{Self-Resolved Diffractive Reactions and Light Cone Wavefunctions}

Diffractive multi-jet production in heavy nuclei provides a novel
way to measure the shape of the LC Fock state wavefunctions and
test color transparency.  For example, consider the
reaction\cite{Bertsch,MillerFrankfurtStrikman,Frankfurt:1999tq}
$\pi A \rightarrow {\rm Jet}_1 + {\rm Jet}_2 + A^\prime$ at high
energy where the nucleus $A^\prime$ is left intact in its ground
state.  The transverse momenta of the jets balance so that $ \vec
k_{\perp i} + \vec k_{\perp 2} = \vec q_\perp < {R^{-1}}_A \ . $
The light-cone longitudinal momentum fractions also need to add to
$x_1+x_2 \sim 1$ so that $\Delta p_L < R^{-1}_A$.  The process can
then occur coherently in the nucleus.  Because of color
transparency, the valence wavefunction of the pion with small
impact separation, will penetrate the nucleus with minimal
interactions, diffracting into jet pairs\cite{Bertsch}. The
$x_1=x$, $x_2=1-x$ dependence of the di-jet distributions will
thus reflect the shape of the pion valence light-cone wavefunction
in $x$; similarly, the $\vec k_{\perp 1}- \vec k_{\perp 2}$
relative transverse momenta of the jets gives key information on
the derivative of the underlying shape of the valence pion
wavefunction\cite{MillerFrankfurtStrikman,
Frankfurt:1999tq,BHDP,Nikolaev:2000sh}.
The diffractive nuclear amplitude extrapolated to $t = 0$ should
be linear in nuclear number $A$ if color transparency is correct.
The integrated diffractive rate should then scale as $A^2/R^2_A
\sim A^{4/3}$ as verified by E791 for 500 GeV incident pions on
nuclear targets\cite{Aitala:2000hc}. The measured momentum
fraction distribution of the jets\cite{Aitala:2000hb} is
consistent with the shape of the pion asymptotic distribution
amplitude, $\phi^{\rm asympt}_\pi (x) = \sqrt 3 f_\pi x(1-x)$.
Data from CLEO\cite{Gronberg:1998fj} for the $\gamma \gamma^*
\rightarrow \pi^0$ transition form factor also favor a form for
the pion distribution amplitude close to the asymptotic solution%
\cite{Lepage:1979zb,Lepage:1980fj} to the perturbative QCD
evolution equation.

The diffractive dissociation of a hadron or nucleus can also occur
via the Coulomb dissociation of a beam particle on an electron
beam (\eg\ at HERA or eRHIC) or on the strong Coulomb field of a
heavy nucleus (\eg\ at RHIC or nuclear collisions at the
LHC)\cite{BHDP}. The amplitude for Coulomb exchange at small
momentum transfer is proportional to the first derivative $\sum_i
e_i {\partial \over \vec k_{T i}} \psi$ of the light-cone
wavefunction, summed over the charged constituents.  The Coulomb
exchange reactions fall off less fast at high transverse momentum
compared to pomeron exchange reactions since the light-cone
wavefunction is effective differentiated twice in two-gluon
exchange reactions. It is also interesting to study diffractive
tri-jet production using proton beams $ p A \rightarrow {\rm
Jet}_1 + {\rm Jet}_2 + {\rm Jet}_3 + A^\prime $ to determine the
fundamental shape of the 3-quark structure of the valence
light-cone wavefunction of the nucleon at small transverse
separation\cite{MillerFrankfurtStrikman}.

There has been an important debate whether diffractive jet
production faithfully measures the light-front wavefunctions of
the projectile. Braun {\em et al.}\cite{Braun:2002wu} and
Chernyak\cite{Chernyak:2002jc} have argued that one should
systematically iterate the gluon exchange kernel from all sources,
including final state interactions.  Thus if the hard momentum
exchange which produces the high transverse momentum di-jets
occurs in the final state, then the $x$ and $k_\perp$
distributions will reflect the gluon exchange kernel, not the
pion's wavefunction. However, it should be noted that the
measurements of pion diffraction by the E791
experiment\cite{Aitala:2000hb} are performed on a platinum target.
Only the part of the pion wavefunction with small impact
separation can give the observed color transparency; {\em i.e.},
additivity of the amplitude on nuclear number.  Thus the nucleus
automatically selects events where the jets are produced at high
transverse momentum in the initial state before the pion reaches
the nucleus\cite{Bertsch}.

The debate\cite{Ivanov:2002hp,Braun:2002wu,Chernyak:2002jc}
concerning the nature of diffractive dijet dissociation also
applies to the simpler analysis of diffractive dissociation via
Coulomb exchange. The one-photon exchange matrix element can be
identified with the spacelike electromagnetic form factor for $\pi
\to q \bar q$; $ \VEV{\pi; P -q | j^+(0)|q \bar q; P}$. Here the
state $\ket{q \bar q}$ is the eigenstate of the QCD Hamiltonian;
it is effectively an `out' state.  If we choose the $q^+=0$ frame
where $q^2 = - \vec q\,^2_\perp$, then the form factor is exactly
the overlap integral in transverse momentum of the pion and $\bar
q q$ LCWFs summed over Fock States. The form factor vanishes at
$Q^2= 0$ because it is the matrix element of the total charge
operator and the pion and jet-jet eigenstates are orthogonal.  The
$n = 2$ contribution to the form factor is the convolution
$\psi_\pi(x,k_\perp-(1-x)q_\perp)$ with $\psi_{\bar q
q}(x,k_\perp)$.  This can be expanded at small $q^2$ in terms of
the transverse momentum derivatives of the pion wavefunction. The
final-state wavefunction represents an outgoing wave of free
quarks with momentum $y, \ell_\perp$ and $1-y, -\ell_\perp$.  To
first approximation the wavefunction $\psi_{\bar q q}(x,k_\perp)$
peaks strongly  at $x = y$ and $k_\perp = \ell_\perp.$ Using this
approximation, the form factor at small $Q^2$ is proportional to
the derivative of the pion light-cone wavefunction $[e_q (1-x)-
e_{\bar q} x] {\partial \over d k_\perp} \psi_\pi(x,k_\perp)$
evaluated at $x = y$ and $k_\perp = \ell_\perp.$ One can also
consider corrections to the final state wavefunction from gluon
exchange.  However, the final quarks are already moving in the
correct direction at zeroth order, so these corrections would be
expected to be of higher order.

\section{Conclusions}

Perturbative QCD provides an important guide to
high momentum
transfer exclusive processes. The theory involves
fundamental
details of hadron structure  at the amplitude level.  The
hadron
wavefunctions required for these perturbative QCD analyses are
also
relevant for computing exclusive heavy hadron decays.

The leading-twist
contributions to exclusive amplitudes derive
from the kinematic regime
where the quarks and gluons propagators
are evaluated in the perturbative
regime.  There are many
successes of the perturbative approach, including
important checks
of color transparency and hadron helicity conservation.
The
successes of perturbative QCD scaling for exclusive processes
at
presently accessible momentum transfers can be understood if
the
effective QCD coupling is approximately constant at the
momentum
transfers scales relevant to the hard scattering amplitudes.
The
Sudakov suppression of the long-distance contributions is
strengthened
if the coupling is frozen because it involves the
exponentiation of a
double logarithmic series.

In this review I have argued that the new
Jefferson Laboratory
measurements of the ratio of proton form factors are
not
necessarily incompatible with the perturbative QCD predictions. I
have
also argued that the apparent discrepancy of theory with the
normalization
of the spacelike pion form factor may be due to the
difficulty of extrapolating electroproduction data from the
off-shell regime to the pion pole.

Further experimental studies, particularly measurements of
electroproduction at Jefferson Laboratory and the study of
two-photon exclusive channels at CLEO and the B-factories have the
potential of providing critical  information on the hadron
wavefunctions as well as testing the dominant dynamical processes
at short distances. Testing quantum chromodynamics to high
precision in exclusive processes is not easy.  Virtually all QCD
processes are complicated by the presence of dynamical higher
twist effects, including power-law suppressed contributions due to
multi-parton correlations, intrinsic transverse momentum, and
finite quark masses.  Many of these effects are inherently
nonperturbative in nature and require detailed knowledge of hadron
wavefunctions themselves.  New systematic approaches to higher
twist contributions are required, such as the recent development
of effective field theories\cite{Bauer:2000yr,Beneke:2002ph}.

Diffractive dijet production on nuclei has provided a compelling
demonstration of color transparency and because of the color
filtering effect of the nuclear target has yielded strong
empirical constraints on the shape of pion distribution amplitude.
I have argued that these ``self-resolving" diffractive processes
can also provide direct experimental information on the light-cone
wavefunctions of the photon and proton in terms of their QCD
degrees of freedom, as well as the composition of nuclei in terms
of their nucleon and mesonic degrees of freedom.

\section{Acknowledgements} I am grateful to Anatoly
Radyushkin and Paul Stoler for their kind invitation to this very
interesting and provocative workshop. I also thank  Volodya Braun,
Carl Carlson, Markus Diehl, Lyonya Frankfurt, Haiyan Gao,  Susan
Gardner, Gudrun Hiller, John Hiller, Paul Hoyer, Dae Sung Hwang,
Peter Kroll, Jerry Miller, Kolya Nilolaev, Stephane Peigne, Mark Strikman,
and Christian Weiss for helpful conversations.

\end{document}